\documentclass{icrc2009}

\usepackage{graphicx}   % for including figures
\usepackage[]{caption}

\usepackage{fixltx2e}
\usepackage{url}

\newcommand{\shorttitle}[1]%
{\markboth{Proceedings of the 31\MakeLowercase{$^{st}$} ICRC, {\L}\'{o}d\'{z} 2009}{#1} }
\newcommand{\etal}{\MakeLowercase{\textit{et al. }}} % "et al."

\begin{document}
\title{Search for neutrinos from transient sources with the ANTARES telescope and optical follow-up observations}

\author{\IEEEauthorblockN{Damien Dornic\IEEEauthorrefmark{1},
			  St\'ephane Basa\IEEEauthorrefmark{2},
			  Jurgen Brunner\IEEEauthorrefmark{1},
			  Imen Al Samarai\IEEEauthorrefmark{1},
			  Jos\'e Busto\IEEEauthorrefmark{1},\\ 			  
			  Alain Klotz\IEEEauthorrefmark{3}\IEEEauthorrefmark{4},
			  St\'ephanie Escoffier\IEEEauthorrefmark{1},
			  Vincent Bertin\IEEEauthorrefmark{1},
			  Bertrand Vallage\IEEEauthorrefmark{5},		  
			  Bruce Gendre\IEEEauthorrefmark{2},\\
			  Alain Mazure\IEEEauthorrefmark{2}and
                          Michel Boer\IEEEauthorrefmark{4} \\ 
			  on behalf the ANTARES and TAROT Collaboration}
                            \\
\IEEEauthorblockA{\IEEEauthorrefmark{1}CPPM, CNRS/IN2P3 - Universit\'e de M\'editerran\'ee, 163 avenue de Luminy, 13288 Marseille Cedex 09, France}
\IEEEauthorblockA{\IEEEauthorrefmark{2}LAM, BP8, Traverse du siphon, 13376 Marseille Cedex 12, France}
\IEEEauthorblockA{\IEEEauthorrefmark{3}OHP, 04870 Saint Michel de l'Observatoire, France}
\IEEEauthorblockA{\IEEEauthorrefmark{4}CESR, Observatiore Midi-Pyr\'en\'ees, CNRS Universit\'e de Toulouse, BP4346, 31028 Toulouse Cedex04, France}
\IEEEauthorblockA{\IEEEauthorrefmark{5}CEA-IRFU, centre de Saclay, 91191 Gif-sur-Yvette, France}
}

% please write the preseter's name and short title (3-4 words maximum)
%    which will appear at the header of the even pages.
\shorttitle{Damien Dornic \etal the TAToO project}
\maketitle

\begin{abstract}
The ANTARES telescope has the opportunity to detect transient neutrino sources, such as gamma-ray bursts, core-collapse 
supernovae, flares of active nuclei... To enhance the sensitivity to these sources, we have developed a new detection method
based on the optical follow-up of "golden" neutrino events such as neutrino doublets coincident in time and space or single
neutrinos of very high energy.
\\
The ANTARES Collaboration has therefore implemented a very fast on-line reconstruction with a good angular resolution. These
characteristics allow to trigger an optical telescope network; since February 2009. ANTARES is sending alert trigger one or
two times per month to the two 25 cm robotic telescope of TAROT. This follow-up of such special events would not only give
access to the nature of the sources but also improves the sensitivity for transient neutrino sources.
\end{abstract}

 
\section{Introduction}
The ANTARES neutrino telescope \cite{BAntares} is located 40 km off shore Toulon, in the South French coast, at about 2500\,m below sea level. 
The complete detector comprises 12 detection lines, each equipped with up to 75 photomultipliers distributed in 25 storeys, which are the sensitive elements. 
Data taking started in 2006 with the operation of the first line of the detector. The construction of the 12 line detector was 
completed in May 2008. The main goal of the experiment is to detect high energy muon induced by neutrino interaction in the 
vicinity of the detector. The detection of these neutrinos would be the only direct proof of hadronic accelerations and so, the discovery of the ultra high 
energy cosmic ray sources without ambiguity.

Among all the possible astrophysical sources, transients offer one of the most promising perspectives for the detection of cosmic 
neutrinos thank to the almost background free search. The fireball model, which is the most commonly assumed, tells us how the GRBs operate but there are still remaining 
important questions such as which processes generate the energetic ultra-relativistic flows or how is the shock acceleration realized. The observation 
of neutrinos in coincidence in time and position with a GRB alert could help to constrain the models.

In this paper, we discuss the different strategies implemented in ANTARES for the transient sources detection. To detect transient 
sources, two different methods can be used \cite{BBasa}.

\section{Transient source detection strategies}
The first one is based on the search for neutrino candidates in conjunction with an 
accurate timing and positional information provided by an external source: the triggered search method. The second one is based 
on the search for high energy or multiplet of neutrino events coming from the same position within a given time window: 
the rolling search method. 

\subsection{The Triggered search}
Classically, GRBs or flare of AGNs are detected by gamma-ray satellites which deliver in real time an alert to the Gamma-ray bursts 
Coordinates Network (GCN \cite{BGCN}). The characteristics (mainly the direction and the time of the detection) of this alert are then distributed to 
the other observatories. The small difference in arrival time and position expected between photons and neutrinos allows a very 
efficient detection by reducing the associated background. This method has been implemented in ANTARES mainly for the GRB 
detection since the end of 2006. Today, the alerts are primarily provided by the Swift \cite{BSwift} and the Fermi \cite{BFermi} satellites. 
\\
Data triggered by more than 500 alerts (including the fake one) have been stored up to now. The "all data to shore" concept used in ANTARES allows to store 
all the data unfiltered during short periods. Based on the time of the external alert, in complement to the standard acquisition strategy, an on-line 
running program stores the data coming from the whole detector during 2 minutes without any filtering. This allows to lower the energy threshold of the event selection 
during the off-line analysis with respect to the standard filtered data. Due to a continuous buffering of data (covering 60s) and thanks to the 
very fast response time of the GCN network, ANTARES is able to recorded data before the detection of the GRB by the satellite \cite{BBouwhuis}. The 
analysis of the data relying on those external alerts is on-going in the ANTARES Collaboration.

Due to the very low background rate, even the detection of a small number of neutrinos correlated with GRBs could set a discovery. 
But, due to the relatively small field of view of the gamma-ray satellites (for example, Swift has a 1.4\,sr field of view), only 
a small fraction of the existing bursts are triggered. Moreover, the choked GRBs without photons counterpart can not be detected 
by this method.

\subsection{The Rolling search}
This second method, originally proposed by Kowalski and Mohr \cite{BKowalski}, consists on the detection of a burst of neutrinos in temporal and 
directional coincidence. Applied to ANTARES \cite{BDornic2}, the detection of a doublet of neutrinos is almost statistically significant. Indeed, 
the number of doublets due to atmospheric neutrino background events is of the order of 0.01 per year when a temporal window of 900\,s and a 
directional one of 3\,$^\circ$ x 3\,$^\circ$ are defined. It is also possible to search for single cosmic neutrino events by 
requiring that the reconstructed muon energy is higher than a given energy threshold (typically above a few tens of TeV). This 
high threshold reduces significantly the atmospheric neutrino background \cite{BDornic}. 
 
In contrary to the current gamma-ray observatories, a neutrino telescope covers instantaneously at least an hemisphere if only up-going events 
are analyzed and even $4\pi$\,sr if down-going events are considered. When the neutrino telescope is running, this method is almost 
100\% efficient. Moreover, this method applies whenever the neutrinos are emitted with respect to the gamma flash. More importantly no assumption is made on the nature 
of the source and the mechanisms occurring inside. The main drawback of the rolling search is that a detection is not automatically 
associated to an astronomical source. To overcome this problem, it is fundamental to organize a complementary follow-up program. The observation of any transient sources will require a quasi real-time analysis and an angular precision lower than a degree.

\section{ANTARES neutrino alerts}
Since the beginning of 2008, ANTARES has implemented an on-line event reconstruction. This analysis strategy contains a very efficient trigger based on local 
clusters of photomultiplier hits and a simple event reconstruction. The two main advantages are a very fast analysis (between 5 and 
10\,ms per event) and an acceptable angular resolution. The minimal condition for an event to be reconstructed is to contain a minimum of 
six storeys triggered on at least two lines. To select a high purity sample of up-going neutrino candidates, one quality cut is 
applied to the result of the $\chi^2$ minimisation of the muon track reconstruction based on the measured time and amplitude of 
the hits. In order to obtain a fast answer, the on-line reconstruction does not use the dynamic reconstructed geometry of the 
detector lines. This has the consequence that the angular resolution is degraded with respect to the one obtained with the standard 
off-line ANTARES reconstruction (of about 0.2 - 0.3\,$^\circ$) which includes the detector positioning.

 \begin{figure}[!t]
  \centering
  \includegraphics[width=2.8in]{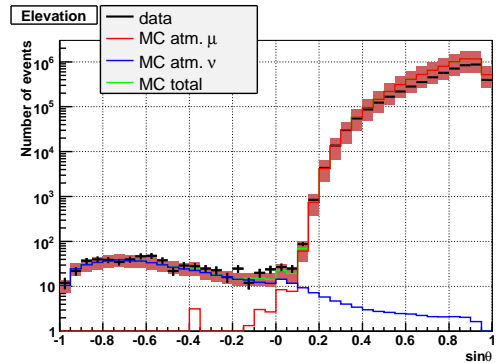}
  \caption{Elevation distribution of the well reconstructed muon tracks (black dots with error bar) recorded in 2008. The blue and red 
    lines represent the Monte-Carlo distribution for atmospheric neutrinos and atmospheric muons (the shaded band contains the systematic error).}
  \label{fig:elev}
 \end{figure}
 
 \begin{figure}[!t]
  \centering
  \includegraphics[width=2.8in]{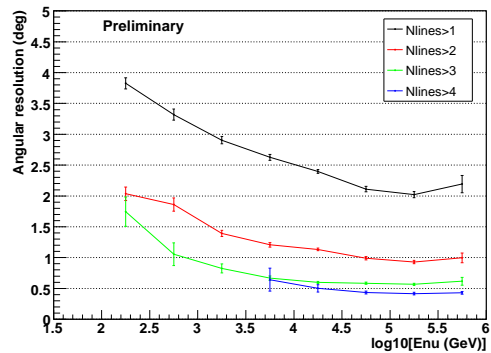}
  \caption{Angular resolution evolution with energy for the event reconstructed on-line with at least 2, 3, 4 and 5 lines.}
  \label{fig:angres}
 \end{figure}
 
In order to set the cuts used for our "golden" neutrino event selection, we have analysed the data taken in 2008 corresponding to 173 active 
days. During this period, around 582 up-going neutrino candidates were recorded. The figure~\ref{fig:elev} shows the elevation distribution of 
the well reconstructed muon events. This plot shows as the same time the distribution of the down-going atmospheric muons and the up-going 
neutrino candidates compare to the distribution predicted by the Monte-Carlo simulations. In order to obtain an angular resolution lower 
than the field of view of the telescope used for the follow-up (around 1\,$^\circ$ in radius), we select reconstructed events which trigger several hits 
on at least 3 lines. The dependence of this resolution with the number of lines used in the fit is shown in the figure~\ref{fig:angres}. For the 
highest energy events, this resolution can be as good as 0.5 degree. An estimation of the energy in the on-line reconstruction is indirectly 
determined by using the number of hits of the event and the total amplitude of these hits. In order to select events with an energy above around 5 TeV, 
a minimum of about 20 storeys and about 180 photoelectrons per track are required. These two different trigger logics applied on the 
2008 data period select around ten events. With a larger delay (few hours after the time of the burst), we are able to run the standard reconstruction tool 
which provides an angular resolution better than around 0.4\,$^\circ$ (still without the dynamic reconstructed geometry of the detector lines).

\subsection{Optical follow-up}
ANTARES is organizing a follow-up program in collaboration with TAROT (\textit{T\'elescope \`a Action Rapide pour les Objets Transitoires}, Rapid 
Action Telescope for Transient Objects, \cite{BTAROT}). This network is composed of two 25\,cm optical robotic telescopes located at Calern (South of France) and La 
Silla (Chile). The main advantages of the TAROT instruments are the large field of view of 1.86\,$^\circ$ x 1.86\,$^\circ$ and their very fast 
positioning time (less than 10\,s). These telescopes are perfectly tailored for such a program. Since 2004, they observe automatically the 
alerts provided by different GRB satellites \cite{BTAROTgrb}. 
\\
As it was said before, the rolling search method is sensitive to all transient sources producing high energy neutrinos. For example, a 
GRB afterglow requires a very fast observation strategy in contrary to a core collapse supernovae for which the optical signal will appear 
several days after the neutrino signal. To be sensitive to all these astrophysical sources, the observational strategy is composed of a real 
time observation followed by few observations during the following month. For each observation, six images integrated over a period of 3 minutes are taken by the telescope. 
We are adapting an image substraction program coming from Supernovae search (\cite{BImage}) in order to look for transient objects in the large field of view. Such a program 
does not require a large observation time. Depending on the neutrino trigger settings, an alert sent to TAROT by this rolling search program would be issued at a 
rate of about one or two times per month.

\subsection{Example of alerts}
\begin{figure}[!t]
  \centering
  \includegraphics[width=2.8in]{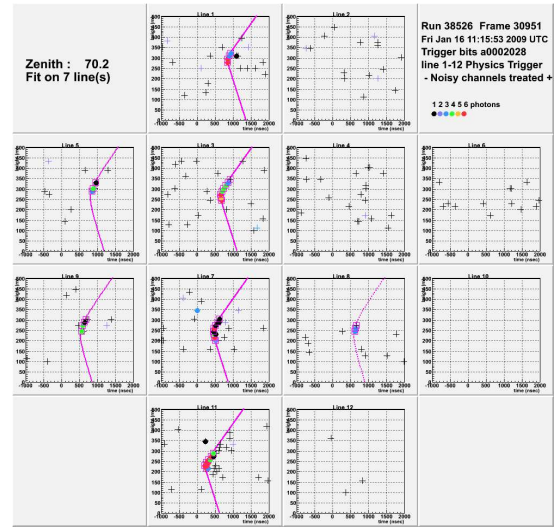}
  \caption{Two dimensional event display of the first neutrino alert reconstructed with 7 lines (not sent to TAROT). Each box represents for each line the height of all the
  hits acquired during a 2 $\mu$s time window versus their times of the detection.}
  \label{fig:event}
 \end{figure}

In the beginning of 2009, with the final run control, 3 alerts were recorded by ANTARES during a test period. Unfortunetly, none of them has been followed by TAROT. 
The figure~\ref{fig:event} shows as example the event display of the first alert recorded in 2009 with ANTARES during this test period. An \textit{a posteriori} search in astronomical
catalogues has shown not interesting objet in a one degree search window correlated with the position of this first alert.

\section{Summary}
The follow-up of golden events would improve significantly the perspective for neutrino detection from transient sources. The most 
important point of the rolling search method is that it is sensitive to any transient source. A confirmation by an optical telescope 
of a neutrino alert will not only give the nature of the source but also allow to increase the precision of the source direction 
determination in order to trigger other observatories (for example very large telescopes for the redshift measurement). The program for 
the follow-up of ANTARES golden neutrino events is operational with the TAROT telescopes since February 2009. It would be also 
interesting to extend this technique to other wavelength observation such as X-ray or radio.


\begin{thebibliography}{99}
\bibitem{BAntares} http://antares.in2p3.fr
\bibitem{BBasa} S. Basa, D. Dornic et al, arXiv:astro-ph/0810.1394
\bibitem{BDornic2} D. Dornic et al, arXiv:astro-ph/0810.1416
\bibitem{BGCN} GCN network, http://gcn.gsfc.nasa.gov
\bibitem{BSwift} S.D. Barthelemy , L.M. Barbier, J.R. Cummings et al, 2005, space Science Reviews, 120, 143
\bibitem{BFermi} http://gammaray.msfc.nasa.gov/gbm
\bibitem{BBouwhuis} M. Bouwhuis, 2005 PhD Thesis, University of Amsterdam
\bibitem{BKowalski} M. Kowalski and A. Mohr, Astroparticle Physics, 27 (2007) 533-538
\bibitem{BDornic} D. Dornic et al, arXiv:astro-ph/0810.1412
\bibitem{BTAROT} M. Bringer, M. Boer et al, 1999, A\&AS, 138 581
\bibitem{BImage} C. Alard, R. Lupton 1998, ApJ, 503, 325
\bibitem{BTAROTgrb} A. Klotz, M. Boer, J.L. Atteia, B. Gendre, AJ 2008 submited

  \end{thebibliography}
\end{document}